\documentclass[aps,prl,twocolumn,groupedaddress]{revtex4-1}
\usepackage{epsfig,amsmath,amssymb,graphics,color,calc}
\usepackage{mathtools}

\newcommand{\re}{\mbox{Re}}

\newcommand{\de}{\textrm{d}}

\begin{document}


\title{Level Statistics and Localization Transitions of L\'evy Matrices}


\author{E. Tarquini\textsuperscript{1,2,3}, G. Biroli\textsuperscript{2}, and M. Tarzia\textsuperscript{1}}

\affiliation{\textsuperscript{1} \mbox{LPTMC,~CNRS-UMR 7600,~Universit\'e~Pierre~et~Marie~Curie,
bo\^ite 121, 75252 Paris c\'edex 05, France}\\\textsuperscript{2} \mbox{Institut de physique th\'eorique, Universit\'e Paris Saclay, CEA, CNRS, F-91191 Gif-sur-Yvette, France}\\ \textsuperscript{3} \mbox{Universit\'e Paris-Sud, 91405-Orsay, France}}



\begin{abstract}
This work provides a thorough study of L\'evy, or heavy-tailed, random matrices (LMs). 
By analyzing the self-consistent equation on the probability distribution of the diagonal elements of the resolvent we
establish the equation determining the localization transition and obtain 
the phase diagram.
Using arguments based on supersymmetric field theory and Dyson Brownian motion we show that 
the eigenvalue statistics is the same one as of the Gaussian orthogonal ensemble in the whole delocalized phase 
and is Poisson in the localized phase. Our numerics confirm these findings, valid in the limit of infinitely large LMs,  
but also reveal that the characteristic scale governing finite size effects diverges much faster than   
a power law approaching the transition and is already very large far from it. 
This leads to a very wide crossover region in which the system looks as if it were in a mixed phase.  
Our results, together with the ones obtained previously, now provide a complete theory of L\'evy matrices. 
\end{abstract}


\maketitle

Since the well-known pioneering
applications of Gaussian random matrices to nuclear spectra, 
random matrix theory (RMT) has found successful applications in many
areas of physics~\cite{metha} and also in other research fields such as  
wireless communications~\cite{telatar}, financial risk~\cite{laloux}, and biology~\cite{bio}.
The reason for such remarkable versatility is that RMT provides {\it universal} results which are independent of 
the specific probability distribution of the random entries: 
only a few features that determine the universality class 
matter.  The most commonly studied RMs 
belong to the Gaussian ensembles~\cite{metha}. They have been analyzed in great depth taking advantage 
of the symmetry under the orthogonal (or unitary/symplectic) group of the probability distribution. 
As an example of universality, $N \times N$ real symmetric RMs,
although they belong to the Gaussian orthogonal ensemble (GOE) only if the elements are Gaussian variables, 
display a GOE-like level statistics also when the distribution of the elements is not Gaussian, provided that it decreases fast enough to infinity~\cite{metha,yau,tao}.\\
There exists, however, a large set of matrices that fall out of the universality classes based on the Gaussian paradigm~\cite{JP}. 
These are obtained when the entries are heavy-tailed i.i.d.~random variables (i.e., with 
infinite variance). The reference case for this different universality class corresponds to entries that are 
{\it L\'evy} distributed. This is the natural generalization of the Gaussian case since the 
limiting distribution of the sum of a large number of heavy-tailed i.i.d. random variables is indeed a 
L\'evy distribution, as is the Gaussian distribution for nonheavy tailed random variables. 
Understanding the statistical spectral properties of these, so called, {\it L\'evy matrices} (LMs) 
is an exciting problem from the mathematical and the physical 
sides~\cite{JP,burda,metz,benarous,giulioJP,benarous2,bordenave,satya}.
They represent a new (and very broad) 
universality class, with different and somehow unexpected 
properties with respect to the Gaussian case.
Actually, a huge variety of distributions in physics and in other disciplines
exhibit power-law behavior. Accordingly, LMs appear in several contexts: in models of spin glasses with dipolar RKKY
interactions~\cite{cizeau}, in disordered electronic systems~\cite{levitov}, in portfolio optimization~\cite{galluccio}, and 
in the study of correlations in big data sets~\cite{politi}, 
just to cite a few.\\
Contrary to the Gaussian case, the theory of random LMs is not yet well established.
LMs were introduced in the pioneering work of Ref.~\cite{JP} and 
further studied in Refs.~\cite{burda,metz,benarous, giulioJP,benarous2,bordenave,satya}. 
By now, 
the behavior of their density of states is well understood 
(even rigorously)~\cite{JP,burda,metz,benarous}. Instead, on finer observables, such as level and eigenfunction statistics, 
there are scarcer and even conflicting results.  
This is probably due to the fact that the behavior of LMs is richer, and hence more difficult to understand, than the one of GOE matrices. 
For instance, a mobility edge separating
high energy localized states from low energy extended states appears within their spectrum~\cite{JP}. It was also argued that they display
a new intermediate mixed phase, characterized by a nonuniversal
level statistics. 
Although some aspects of the scenario put forward in Ref.~\cite{JP} are in contradiction with recent rigorous results~\cite{bordenave},
such mixed phase could indeed exist and actually be related to the one recently observed in the Anderson model on the
Bethe lattice~\cite{noi,scardicchio}.  It may be the simplest case of the  nonergodic delocalized phase  advocated for quantum many body disordered systems in Ref.~\cite{altshuler}. 

\begin{figure}
\includegraphics[width=0.42\textwidth]{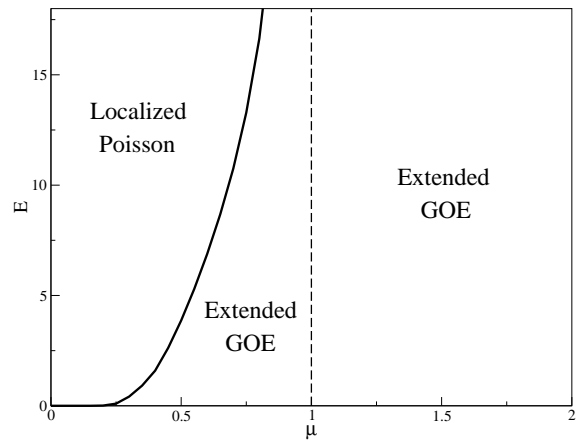}%
\caption{\label{fig:pd}
Phase diagram of LMs in the $\mu$-$E$ plane.} 
\end{figure}

In the following we focus on $N \times N$ real symmetric matrices $\mathcal{H}$ with entries $h_{ij} = h_{ji}$
distributed independently according to a law, $P(h_{ij}) =N^{1/\mu} f(N^{1/\mu} h_{ij})$, 
characterized by heavy tails: 
\[
P(h_{ij})\simeq \frac{\mu}{2 N |h_{ij}|^{1+\mu}} \,\,,\,\, |h_{ij}|\rightarrow \infty \,\,;\,\,\mu<2\,.
\]
The specific form of $f(x)$ does not matter. 
For concreteness in numerical applications we will focus on a Student distribution with exponent $1+\mu$ and symmetric entries, $f(x)=f(-x)$. 
The scaling of the entries with $N$ is such that almost all eigenvalues are $O(1)$ for $N\rightarrow \infty$.\\
The first issue we address is determining the localization-delocalization transition line $E^\star(\mu)$ in the $E$-$\mu$ plane. 
In order to do so, we focus on the statistics of the diagonal elements of the resolvent matrix $\hat{G} = [(E - i \eta) \mathcal{I} - \mathcal{H}]^{-1}$,
which allows one to compute in the $\eta \to 0^+$ limit spectral properties of $\mathcal{H}$ such as the global density of states 
$\rho(E) = (1/N) \sum_{n=1}^N \delta(E - \lambda_n) = \lim_{\eta \to 0^+} (1/N \pi) \sum_{i=1}^N \Im G_{ii}$
and the average inverse participation ratio (IPR) $\langle \Upsilon_{2,n} \rangle = \langle \sum_{i=1}^N | \langle{i}|{n}\rangle |^4 \rangle = \lim_{\eta \to 0^+} (1/N) \sum_{i=1}^N \eta |G_{ii}|^2$.
As shown in Refs.~\cite{JP, burda,metz,benarous}, the probability distribution $Q(G)$ of a given $G_{ii}$ is obtained in the large-$N$ limit from the equation:
\begin{equation}
\label{eq:G}
G_{ii}^{-1} \overset{d}{=} E - i \eta - \sum_{j=1}^N h_{ij}^2 G_{jj} \, ,
\end{equation}
where all correlations between the terms on the rhs can be neglected
and $\overset{d}{=}$ denotes the equality in distribution between random variables. 
This leads to a self-consistent equation on $Q(G)$,
whose analysis yields 
the results on the density of states 
obtained in Refs.~\cite{JP,benarous}: 
For $\mu<2$, $\rho (E)$ is a $\mu$-dependent 
symmetric distribution with support on the whole real axis and fat tails with exponent $1+\mu$ 
(the semicircle law is recovered for $\mu > 2$ only). 
There are several complementary ways to obtain the localization transition from the statistics of the $G_{ii}$s. 
We have followed the one more likely to receive a rigorous treatment, as it was shown for the Anderson transition on the Bethe lattice~\cite{warzel}. 
It consists in studying the stability of the localized phase, checking whether adding a small imaginary part to $G_{ii}$ is an unstable perturbation~\cite{abou}. 
Such stability is governed by an eigenvalue equation for the same integral
operator found in Ref.~\cite{JP}, whose analysis can be considerably simplified,
as shown in Ref.~\cite{EPAPS}, and boils down to
the following closed equation for the mobility edge $E^\star (\mu)$, which is one of the main results of this work: 
\begin{equation}~\label{eq:mobility} 
K_{\mu}^2 \left( s_\mu^2 - s_{1/2}^2 \right) |\ell (E^\star)|^2 - 2 s_\mu K_{\mu} 
\, \Re \ell (E^\star) + 1 = 0 \, ,
\end{equation}
where $K_{\mu} = \mu \, \Gamma( 1/2 - \mu/2)^2 /2$, $s_\mu = \sin(\pi \mu/2)$ and
$\ell (E) = \int_0^{+\infty} k^{\mu - 1} \, \hat{L}_{\mu/2}^{C(E),\beta(E)} (k) \, e^{i k E} \, \textrm{d}k/\pi$. The function 
$\hat{L}_{\mu/2}^{C(E),\beta(E)} (k)$ is the Fourier transform of the probability distribution of the real part of the self-energy, 
that previous works have shown to be a L\'evy stable distribution with exponent $1+\mu/2$ and parameters $C(E)$ and $\beta(E)$ determined self-consistently~\cite{JP,burda,benarous}. 
This equation has a solution for $\mu \in (0,1)$ only. For $\mu \rightarrow 1$
we find that $E^\star (\mu)$ diverges as 
$(1-\mu)^{-1}$. In Fig.~\ref{fig:pd} we show the numerical solution of Eq.~(\ref{eq:mobility}) for several values of $\mu$ (we only consider $E>0$ since the spectral properties are symmetric around zero). 
This quantitative phase diagram is in agreement with the sketch of Ref.~\cite{JP} and the numerics 
of Ref.~\cite{metz} (except for $\mu>1$ where the results were likely inaccurate due to the very large values of $E$ that had to be explored).  \\
We now address more subtle issues related to the level and eigenfunction statistics. 
We present first, two analytical arguments which show that the statistics is a GOE 
in the whole delocalized phase and Poisson in the localized phase for $N\rightarrow \infty$.
The former is based on the supersymmetric zero-dimensional field theory introduced for random GOE matrices~\cite{efetov}. 
Since we follow closely the techniques developed in Refs.~\cite{mirlinsigma,mirlinbethe} we just discuss the main steps and refer to Ref.~\cite{EPAPS}  and a longer paper~\cite{long} for more details. 
The starting point is the field theory $Z=\int \prod_i \textrm{d}\Phi_i e^{S[\Phi_i]}$, with the action
\[
S=\frac i 2 \left(\sum_{l,m} \Phi_l^\dagger \mathcal{L} (E\delta_{lm}-h_{lm})\Phi_m +\sum_l \Phi_l^\dagger \Phi_l \, \frac{r+i0^+}{N} \right) \, .
\] 
The field $\Phi_i$ is a eight-component super-vector $(\Phi^{(1)}_i,\Phi^{(2)}_i)=
(S_i^{a},S_i^{b},\chi_i,\chi_i^*,P_i^{a},P_i^{b},\eta_i,\eta_i^*)$, where each of the four component supervector $\Phi^{(1,2)}_i$
is formed by two real and two Grassman variables. The matrix $\mathcal{L}$ is diagonal with elements $(1,1,1,1,-1,-1,-1,-1)$. 
The level statistics, in particular the density of states and the correlation between two levels at distance $r/2N$, can be obtained 
from correlation functions of the fields~\cite{efetov}.  
Averaging over the 
the matrix elements and introducing the function $\rho(\Phi)=\frac 1 N \sum_i \delta(\Phi-\Phi_i)$ one can rewrite $Z$ as 
$\int \! {\mathcal D} \rho(\Phi) e^{S[ \rho]}$ with the action reading
\begin{displaymath}
\begin{split}
&S=\frac i 2 NE \!  \int \! \textrm{d} \Phi \rho(\Phi)   \Phi^\dagger \mathcal{L} \Phi + \frac i 2 (r+i0^+) \! \int \! \textrm{d}\Phi   \rho(\Phi) \Phi^\dagger \Phi \\
&-N \!\! \int\!\! \textrm{d}\Phi  \rho(\Phi)\log  \rho(\Phi)+\frac i 2N \!\!\int\!\! \textrm{d} \Phi \textrm{d}\Psi  \rho(\Phi) C(\Phi^\dagger \mathcal{L} 
\Psi)\rho(\Psi)\, ,
\end{split}
\end{displaymath}
where $C(y)=\mu \int \frac{\textrm{d}x}{2 |x|^{1+\mu}}[\exp(-ixy)-1]$. 
Since the second term is subleading compared to the other three that are $O(N)$, 
one can neglect it at first and perform a saddle point. The solution of the corresponding equation reads
\[
\rho(\Phi)\!=\!\int \! d \Sigma R(\Sigma) \exp\left(\frac i 2 \Phi^\dagger \mathcal{L} \Phi (E-\Re \Sigma)+\frac 1 2 \Phi^\dagger \Phi \Im \Sigma
\right) \, ,
\] 
where, as it can be shown in full generality \cite{mirlinsigma,EPAPS}, $R(\Sigma)$ is the probability distribution of the local self-energy, 
which coincides with the complex L\'evy stable law rigorously proven in Ref.~\cite{benarous} (see Ref.~\cite{EPAPS}).
Note that the saddle point equation is invariant under the symmetry $\Phi\rightarrow \mathcal{T} \Phi$ where the super matrix $\mathcal{T}$ 
verifies the equation 
$\mathcal{T}^\dagger \mathcal{L} \mathcal{T} = 1$. Thus given a solution $\rho(\Phi)$, $\rho_{\mathcal{T}} (\Phi)=\rho(\mathcal{T} \Phi)$ is also a solution. 
The localization transition corresponds to the breaking of this symmetry~\cite{efetov,mirlinbethe}: 
in the localized phase the typical value of the imaginary part of the self-energy is zero, whereas it is finite in the delocalized phase. 
In consequence, in the former case $\rho(\Phi)$ is a function of $\Phi^\dagger \mathcal{L} \Phi$ only, invariant under the symmetry generated 
by $\mathcal{T}$,  
whereas in the latter it depends also on $\Phi^\dagger \Phi$. Since this dependence breaks the symmetry there 
is a manifold of solutions  $\rho_{\mathcal{T}} (\Phi)$. It is the integration over this manifold that leads to 
GOE statistics for the level correlations. The derivation is identical to the one presented in Ref.~\cite{mirlinbethe} 
since the only term in the action that depends on $\mathcal{T}$, i.e.,~that breaks the symmetry, 
is the $r$ one as it happens for Erd\"os-R\'enyi graphs~\cite{mirliner} and GOE RMs~\cite{efetov}. 
In the localized phase, the saddle point solution is instead unique. Therefore no integration over $\mathcal{T}$ has to be performed and this leads to uncorrelated levels, i.e.,~Poisson statistics~\cite{mirlinbethe}.\\
Let us now turn to the other analytical argument, which is very straightforward but limited to $\mu>1$ only. 
Taking inspiration from the recent mathematical breakthrough on RMT~\cite{yau}, we slightly modify 
the distribution $P(h_{ij})$ into $(1-\epsilon) P(h_{ij})+\epsilon N^{1/\mu}W(N^{1/\mu}h_{ij})$
where $W(x)$ is a Gaussian distribution with unit variance. This is equivalent to modifying $\mathcal{H}$ into $\mathcal{H}_{\epsilon}=(1-\epsilon) \mathcal{H}+\epsilon \mathcal{W}$
where $\mathcal{H}$ is a LM and $\mathcal{W}$ a very small GOE matrix
whose elements have exactly the same scaling with $N$ than the ones of $\mathcal{H}$. Since this change does not alter the fat tails of the matrix elements, 
one naturally expects $\mathcal{H}_\epsilon$ and $\mathcal{H}$ to be in the same universality class for any $\epsilon<1$ and in 
particular for $\epsilon \rightarrow 0$. The statistics of the modified LM\,---\,and, by the previous argument, of $\mathcal{H}$\,---\,can 
be obtained using the Dyson Brownian motion (DBM): $\mathcal{H}_\epsilon$ can be interpreted, in the basis that diagonalizes $\mathcal{H}$,  as 
a diagonal matrix to which an infinite number of infinitesimal GOE matrices
have been added. The probability of the eigenvalues 
of $\mathcal{H}_\epsilon$ is therefore given by the DBM starting from the eigenvalues of $\mathcal{H}$, and evolving over a fictive time of the order $N^{-1/\mu}$. 
Recent rigorous results~\cite{yau} guarantee that the DBM has enough ``time'' to reach its stationary distribution, 
which is the GOE distribution, if $N^{-1/\mu}\gg N^{-1}$ and the typical level spacing of $\mathcal{H}$ is $O(1/N)$\,---\,a very reasonable assumption that agrees well with the numerics. 
This implies that for $\mu>1$ the level statistics of the modified LM, and hence of the original LM too, is indeed GOE-like in the bulk of the spectrum~\cite{footnote-edge} (see Refs.~\cite{EPAPS,long} for more details). 
\begin{figure}
\includegraphics[width=0.42\textwidth]{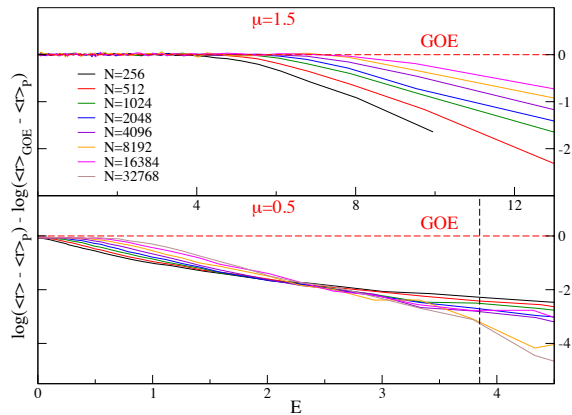}%
\caption{\label{fig:r-q}
$\ln[(\langle r \rangle - \langle r \rangle_P)/(\langle r \rangle_{\rm {GOE}} - \langle r \rangle_P)]$ as a function of $E$ for different system sizes and for $\mu=1.5$ (top) and $\mu=0.5$ (bottom).
The 
dashed line represents the position of the mobility edge, $E^\star \simeq 3.85$.}
\end{figure}

We now present several numerical results with the aim of backing up our previous analytical arguments and also of studying 
the behavior of large {\it but finite} LMs.  In applications $N$ is never truly infinite, actually in several 
cases it can be just a few thousand. Thus, it is of paramount importance to study finite size effects and determining the
characteristic value of $N$ above which the $N\rightarrow \infty$ limit is recovered.  
We performed exact diagonalization of LMs 
 for several system sizes $N=2^n$, from $n=8$ to $n=15$ and averaging 
over $2^{22-n}$ realizations of the disorder.
We have resolved the
energy spectrum in $64$ small intervals $\nu$, 
centered around the energies $E_\nu = \langle \lambda_n \rangle_{n \in \nu}$, 
and analyzed the statistics of eigenvalues and eigenfunctions in each one of them. 
We have focused on several observables that display 
different universal behaviors in the GOE and Poisson regimes: 
The first probe, introduced in Ref.~\cite{huse}, is the ratio of adjacent gaps 
$r_n = \min \{ \delta_n, \delta_{n+1} \} / \max \{ \delta_n, \delta_{n+1} \}$
where $\delta_n = \lambda_{n+1} - \lambda_n \ge 0$ denotes the level spacings between 
neighboring eigenvalues. It has different universal distributions in the GOE and Poisson
cases encoding, respectively, the repulsion or the independence of levels. 
The second one is the overlap between eigenvectors corresponding to 
subsequent eigenvalues, defined as $q_{n} = \sum_{i=1}^N | \langle{i}|{n}\rangle |
| \langle{i}|{n+1}\rangle |$. Its typical value $q_\nu^{\rm {typ}} = e^{\langle \ln q_n \rangle_{n \in \nu}}$
allows us to make the difference between the localized phase, in which 
subsequent eigenvectors do not overlap ($q^{\rm {typ}}=0$), and the 
delocalized GOE one in which they do ($q^{\rm {typ}} = 2/\pi$). 
Finally,  the wave function support set, recently introduced
in Ref.~\cite{scardicchiosupport}, is defined for an eigenvector ${n}$ with sites ordered 
according to $|\langle{i}|{n}\rangle| > |\langle{i+1}|{n}\rangle|$ as the sets of sites $i < S_\epsilon^{(n)}$ such that
$\sum_{i=1}^{S_\epsilon^{(n)}} |\langle{i}|{n}\rangle|^2 \le 1 - \epsilon < \sum_{i=1}^{S_\epsilon^{(n)}+1} |\langle{i}|{n}\rangle|^2$.
The scaling 
of $S_\epsilon^{(n)}(N)$ for $N \to \infty$ and $\epsilon$ arbitrary small but finite allows us to
discriminate between a localized and extended phase. 
\begin{figure}
\includegraphics[width=0.42\textwidth]{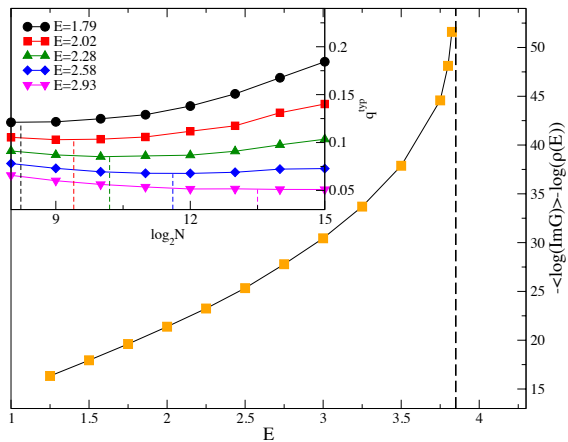}%
\caption{\label{fig:length} Main panel: $\log N_m' (E)= - \log (\Im G_{ii}^{\rm {typ}} \rho(E))
= - \langle \log \Im G_{ii} \rangle - \log (\langle \Im G \rangle / \pi)$ 
as a function of $E$ for $\mu=0.5$. Inset: $q_{\rm {typ}}$ as a function of $N$ for different energies
and $\mu=0.5$, showing the position of $N_m(E)$.}
\end{figure}
The analysis of all these probes clearly shows that for $\mu>1$ the eigenvalues and eigenvectors statistics is a GOE in the limit of large $N$, 
in agreement with our previous arguments (and also with the rigorous results on the delocalized nature of
wave functions~\cite{bordenave}). As an example we show in the top panel of Fig.~\ref{fig:r-q} 
the behavior of the average value of $r_n$ for $\mu=1.5$. It
clearly converges in the limit $N\rightarrow \infty$ and for any energy $E$ to the value $\langle r_n \rangle_{\rm {GOE}}\simeq 0.53$ 
characteristic of GOE statistics. 
All other probes show a similar convergence to the values expected for GOE statistics (see Ref.~\cite{EPAPS} for 
the corresponding plots).    
This is no longer true for $\mu < 1$, where the situation is more involved. 
In the lower panel of Fig.~\ref{fig:r-q} we show again the behavior of the average of  $r_n$ but now for $\mu=0.5$. For small and large energies we find the values 
$\langle r \rangle_{\rm {GOE}} \simeq 0.53$ and $\langle r \rangle_{P} \simeq 0.39$ corresponding respectively to 
GOE and Poisson statistics. Moreover, the curves corresponding to different values of $N$ seem to cross much before the localization transition, that our previous analytical results located at 
$E^\star \simeq 3.85$ for $\mu=0.5$. 
If this were representative of the truly asymptotic large-$N$ behavior then it would possibly signal  
the existence of a mixed phase which could be {\it delocalized but nonergodic}, i.e.,~not displaying GOE statistics.
However, analyzing carefully  the data---thanks to the large number of samples used to average over the disorder---we find that the crossing point is in fact very slowly drifting towards higher energies as $N$ is increased. 
The same behavior is found for {\it all} the probes we studied. As an example, 
in the inset of Fig.~\ref{fig:length} we plot $q_\nu^{\rm {typ}}$ as a function
of the system size for energies belonging to the crossing region. This indeed shows that 
$q_{\rm {typ}}$ is a nonmonotonic function of $N$.
We can then define a characteristic matrix size, $N_m(E)$, such that
for $N \ll N_m(E)$ the statistics appears to be intermediate between Poisson and GOE (see Ref.~\cite{EPAPS}), whereas for $N \gg N_m(E)$ it tends again toward 
GOE.\\ 
The existence of a crossover size can be understood from the properties of the distribution $Q(G)$. 
What characterizes the delocalized phase is that, at any site $i$, the imaginary part of $G_{ii}$ receives an infinitesimal contribution from 
an infinite number of eigenfunctions.  
This leads to a typical value of $G_{ii}$ (defined as $\Im G_{ii}^{\rm {typ}} = e^{\langle \log \Im G_{ii} \rangle}$) which is finite for $N\rightarrow \infty$ and $\eta \rightarrow 0$. 
Instead, $\Im G_{ii}^{\rm {typ}} = 0$ in the localized phase. 
Approaching the transition from the delocalized side, $\Im G_{ii}^{\rm {typ}}$
becomes extremely small. Thus, one needs to take large enough systems 
in order to realize that it is different from zero, and hence that the system is in the delocalized and GOE-like phase.
The argument, which is based on the interpretation of 
$\Im G_{ii}$ as the local density of states, is as follows. 
 The number of states per unit of energy close to $E$ is $N\rho(E)$.
This number, multiplied
by the typical value of the local density of states, has to be larger than one in order to be in a regime representative
of the large-$N$ limit. This defines the crossover scale $N_m'(E)\propto 1/(\Im G_{ii}^{\rm {typ}} \rho(E))$. 
We have compared numerically $\ln N_m^\prime (E)$ and $\ln N_m(E)$ and found that they are indeed proportional (see Ref.~\cite{EPAPS} 
for a plot), thus showing that our argument correctly captures the origin of the finite size effects.
We plot the crossover scale [actually $N_m^\prime(E)$] as a function of $E$ in Fig.~\ref{fig:length} for $\mu=0.5$: 
it diverges very fast approaching $E^\star (\mu)$. A good fit is provided by an essential singularity.   
These results therefore unveil what is the mechanism responsible for the non-GOE statistics 
observed for finite LMs in a wide regime before the localization transition.

In conclusion, we have presented a thorough analysis of the eigenvalues and eigenvectors statistics of
random L\'evy matrices. 
We have shown that the
localization and the level statistics transitions coincide but also unveil the existence of a 
crossover scale which is very large even far from the transition. 
Thus, many practical cases are expected to be in the $N\ll N_m(E)$ regime. 
In consequence, the mixed behavior proposed in Ref.~\cite{JP} will be often present 
in practice even though it is absent in the large-$N$ limit. 
Our work, together with the results obtained previously, now provides a complete theory of LMs. \\
There are several directions worth pursuing more. It would be interesting to determine analytically the 
form of the divergence of $N_m(E)$. On the basis of our numerics and in analogy with previous works~\cite{efetov,mirliner,mirlinbethe} we expect $N_m(E)\propto e^{c/(E^\star-E)^a}$. 
Most probably, the emergence of the crossover scale
producing an apparent mixed phase takes place in several other related situations (e.g., Ref.~\cite{noi}) that are, therefore, to be reanalyzed. 
Finally, our results provide a guideline for mathematicians working on RMT. Thanks to the 
recent advances in the mathematical analysis of random matrices~\cite{yau} and localization phenomena~\cite{warzel}
our findings are likely to be rigorously proven in a not too distant future.

\begin{acknowledgments}
We thank G. Ben Arous, J.-P. Bouchaud, L. Erd\"os, and S. Warzel for helpful discussions and
acknowledge support from the ERC grant NPRGGLASS. Part of this work was done at the ESI. 
\end{acknowledgments}


\onecolumngrid

\newpage

\begin{center}
\Large{\textbf{EPAPS}}
\end{center}

In this supplementary material we provide more details and results related to several points discussed in the main text.
For definiteness we shall consider $N \times N$ real symmetric (Wigner)
L\'evy Matrices $\mathcal{H}$ with entries $h_{ij} = h_{ji}$
distributed independently according to a student distribution with exponent $1+\mu$ and typical value of order
$N^{-1/\mu}$:
\begin{equation} \label{eq:Ph}
P(h_{ij}) = \theta \left( |h_{ij}| > N^{-1 / \mu} \right) \frac{\mu}{2 N |h_{ij}|^{1+\mu}} \, .
\end{equation}

\section{Computation of the mobility edge}

\subsection{The linearized recursive equations}

In order to determine the position of the mobility edge, it is convenient 
to introduce the self-energy $\Sigma_{ii} = E - i \eta - G_{ii}^{-1} = S_i + i \Delta_i$,
and linearize the recursive equations for the diagonal elements of the resolvent matrix
[Eq.~(1) of the main text] with respect to their imaginary part
in the limit $\eta \to 0^+$: 
\begin{subequations}
\begin{align} \label{eq:S}
S_i & \overset{\rm d}{=} \sum_{j=1}^N h_{ij}^2 \, \Re G_{jj} = 
\sum_{j=1}^N \frac{h_{ij}^2 (E - S_j)}{(E - S_j)^2 + (\eta + \Delta)^2} \simeq 
\sum_{j=1}^N \frac{h_{ij}^2}{E - S_j} \, , \\
\label{eq:Delta}
\Delta_i & \overset{\rm d}{=} \sum_{j=1}^N h_{ij}^2 \, \Im G_{jj} = 
\sum_{j=1}^N \frac{h_{ij}^2 (\eta + \Delta_j)}{(E - S_j)^2 + (\eta + \Delta)^2} \simeq 
\sum_{j=1}^N \frac{h_{ij}^2}{(E - S_j)^2} \, \Delta_j \, .
\end{align}
\end{subequations}
The recursion relation~(\ref{eq:S}) for the real part of the self-energy
totally decouples from the equation~(\ref{eq:Delta}) for its imaginary part.
The marginal probability distribution of the real part can thus be computed  
using the generalized central limit theorem.
One possible route to obtain the recursion relations [Eq.~(1) of the main text and Eqs.~(\ref{eq:S}) and  (\ref{eq:Delta})]
is provided by the cavity method~\cite{JP-SM}. Within this framework, the $G_{jj}$s appearing in Eq.~(\ref{eq:S})
are the diagonal element of the resolvent of a LM of $(N-1)\times(N-1)$ elements, where the $i$-th row and column
have been removed. As a consequence in the $N \to \infty$ limit the matrix elements $h_{ij}$ and the terms $G_{jj}$ are 
independent and uncorrelated by construction.
Since the variance 
of the entries is infinite for $\mu < 2$, by virtue of the generalized central limit theorem 
the probability distribution of the variable $S$ tends, for $N \to \infty$,
to a L\'evy stable distribution, $L_{\mu/2}^{C(E),\beta(E)} (S)$, with
stability index $\mu/2$, and ``effective range'' $C(E)$ and ``asymmetry parameter''
$\beta(E)$ given by:
\begin{equation} \label{eq:C-beta-EPAPS}
\left \{
\begin{split} 
C & =  \Gamma \left(1 - \frac{\mu}{2} \right) \cos \left(
\frac{\pi \mu}{4} \right) \, \frac{1}{N} \sum_{j=1}^N \left | \Re G_{jj} \right |^{\frac{\mu}{2}} \, , \\
\beta & =  \frac{ \frac{1}{N} 
\sum_{j=1}^N \left | \Re G_{jj} \right |^{\frac{\mu}{2}} \textrm{sign} 
(\Re G_{jj} )}{
\frac{1}{N} 
\sum_{j=1}^N \left | \Re G_{jj} \right |^{\frac{\mu}{2}}} \, .
\end{split}
\right .
\end{equation}
In the $N \to \infty$ limit the diagonal elements of the resolvent 
become independent and identically distributed. We can therefore replace the sums in the
r.h.s.~of Eqs.~(\ref{eq:C-beta-EPAPS}) by integrals over the marginal probability
distribution $\tilde{Q}_R (\Re G)$.
In the limit of vanishing imaginary part, one has that $S_i = E - 1/\Re G_{ii}$. Hence by changing variable one has: 
\begin{displaymath}
\tilde{Q}_R (\Re G) \, \textrm{d} \Re G
= L_{\mu/2}^{C(E),\beta(E)} \left( E - \frac{1}{\Re G} \right) \frac{\textrm{d} \Re G }{ |\Re G|^2} \, .
\end{displaymath}
As a consequence, Eq.~(\ref{eq:C-beta-EPAPS}) can be written as a set of two coupled
self-consistent equation for the parameters $C(E)$ and $\beta(E)$~\cite{JP-SM,benarous-SM}, which can be solved numerically with a reasonably high degree of accuracy.
\begin{subequations}
\begin{align} \label{eq:C-self-EPAPS}
C(E) & = \Gamma \left(1 - \frac{\mu}{2} \right) \cos \left(
\frac{\pi \mu}{4} \right) \int_{-\infty}^{+\infty} L_{\mu/2}^{C(E),\beta(E)} (S) \,
\left| E - S \right|^{-\frac{\mu}{2}} \textrm{d} S \, ,\\
\label{eq:beta-self-EPAPS}
\beta(E) & = \frac{\int_{-\infty}^{+\infty} L_{\mu/2}^{C(E),\beta(E)}  (S) \, \,
\textrm{sign} \left( E - S \right) \left| E - S \right|^{-\frac{\mu}{2}} \textrm{d}S }
{\int_{-\infty}^{+\infty} L_{\mu/2}^{C(E),\beta(E)} (S) \,
\left| E - S \right|^{-\frac{\mu}{2}} \textrm{d}S} \, .
\end{align}
\end{subequations}

\subsection{The mapping to directed polymers in random media}

We now focus on the recursive equation~(\ref{eq:Delta}) for the imaginary part of the self-energy.
If one replaces the $\Delta_j$s appearing in the
sum of the r.h.s.~by their expression in terms of their ``neighbors'' iteratively, say $R$ times, one obtains:
\begin{displaymath} 
\Delta_{i_1} = \sum_{i_2 \neq i_1} \frac{h_{i_1 i_2}^2}{(E - S_{i_2})^2} \sum_{i_3 \neq i_2} \frac{h_{i_2 i_3}^2}{(E - S_{i_3})^2}
\cdots \sum_{i_R \neq i_{R-1}} \frac{h_{i_{R-1} i_R}^2}{(E - S_{i_R})^2} \, \Delta_{i_R} \, .
\end{displaymath}
As a result, the equation for the imaginary part of the self-energy 
can be interpreted as a sum over directed paths of length $R$ originating from the site $i_1$.
For each edge $(i_{n+1},i_n)$ crossed by a given path, the contribution to $\Delta_{i_1}$ coming from the path 
picks the random factor $h_{i_n i_{n+1}}^2/(E - S_{i_{n+1}})^2$. 
The equation above can be then telescoped as:
\begin{equation} \label{eq:DPRM-EPAPS}
\Delta_{i_1} = \sum_{\mathcal{P}} \left( \prod_{(i_{n+1},i_n)} \frac{h_{i_n,i_{n+1}}^2}
{(E - S_{i_{n+1}})^2} \right) \Delta_{i_R} \, ,
\end{equation}
The imaginary part of the self-energy thus 
satisfies the exact same recursive equation as the partition function
of directed polymers in random media (DPRM)~\cite{derrida} in presence of quenched (correlated) bond disorder
$e^{-\epsilon_{ij}} = h_{ij}^2/(E - S_j)^2$.
One can therefore use the results established in this context to analyze the properties of the distribution of 
$\Im \Sigma$ and study the localization transition of LMs, as detailed below.

It is well known that, depending on the strength of the disorder, DPs have a freezing transition 
(called one-step Replica Symmetry Breaking)
akin to the glass phase of the Random Energy Model~\cite{derrida,derrida1}.
The sum in~(\ref{eq:DPRM-EPAPS}) is over an exponential number of paths, $(N-1)!/(N-R-1)! \sim N^R$. 
In the large $R$ limit, two
cases are possible: the sum is dominated either by few paths that give a $O(1)$ contribution, or by an exponential
number of paths, each of them giving a very small contribution but such that their sum is $O(1)$.
The freezing-glass transition of DPRM corresponds to the transition between these two regimes.
It is then clear that in fact, {\it mutatis mutandis}, the glass phase
of DPRM corresponds to the Anderson localized regime, where the imaginary part of the self-energy
is of order $\eta$ with probability $1$ on (almost) all the sites, except on extremely rare and distant resonances
where it is of $O(1)$ and the sum in~(\ref{eq:DPRM-EPAPS}) is dominated by very few paths. 
Conversely, the ergodic phase of DPs corresponds to the delocalized regime, where
$\Delta_i$ is finite on all the sites and all paths give a non-zero contribution to the sum.

Such transition is related to an ergodicity breaking. 
In order to determine the transition point, let us imagine to introduce $n$ replicas of the system with the same 
realization of the disorder (i.e., $n$ identical LMs with the same matrix elements).
Following the analogy with DPRM, the ``partition function'' of the replicated system is
\begin{displaymath}
Z^n = \left[ \sum_{\mathcal{P}} \prod_{(i_{n+1},i_n)} e^{-\epsilon_{ij}} \right ]^n \, ,
\end{displaymath}
and (minus) the (quenched) ``free energy'' per site is $\phi = \overline{\log Z^n}/Rn$.
(In the following for simplicity we will instead compute the ``annealed'' free energy $\phi = \log \overline{Z^n}/Rn$, 
since it yields the same transition point as the quenched one).
In the glass phase the sum in~(\ref{eq:DPRM-EPAPS}) is dominated by few, $O(1)$, paths. 
One can thus assume that the $n$ replicas are 
divided in $n/m$ groups of $m$ replicas all freezed in the same specific path (one-step Replica Symmetry Breaking ansatz).
One than has:
\begin{displaymath}
Z^n = \left[ \sum_{\mathcal{P}} \prod_{(i_{n+1},i_n)} e^{-m \epsilon_{ij}} \right ]^{\frac{n}{m}} \, .
\end{displaymath}
In this case the (annealed) replicated free energy of DPRM reads:
\begin{equation} \label{eq:phi-EPAPS}
\phi(m,E) = \frac{1}{Rm} \log \overline{ \left( 
\sum_{\mathcal{P}} \prod_{(i_{n+1},i_n)} \left| \frac{h_{i_n,i_{n+1}}}{E - S_{i_{n+1}}} \right|^{2m}
\right)} \, ,
\end{equation}
which gives the behavior of the typical value of $\Delta_i^m$.
The free energy needs then to be extremized with respect to the parameter $m$: 
$\partial \phi / \partial m |_{m = m^\star} = 0$. 
If one finds that $\phi(m^\star,E)<0$, then this implies that the partition function is exponentially small and that the typical value of the 
imaginary part of the self-energy vanishes exponentially under iteration. Hence the system is localized.
Conversely, for $\phi(m^\star,E)>0$ the typical value of the imaginary
part of the self-energy grows exponentially under iteration and the system is in the delocalized phase.
The Anderson localization is therefore given by:
\begin{displaymath}
\left \{
\begin{split}
& \left . \frac{\partial \phi (m,E)}{\partial m} \right |_{m = m^\star} \!\! = 0 \, , \\
& \phi(m^\star,E)=0 \, .
\end{split}
\right .
\end{displaymath}
For the Anderson model on the Bethe lattice the localization transition takes place for $m=1/2$,
as it has been rigorously proven in~\cite{warzel-SM,bapst} and indirectly found in~\cite{abou-SM}.
It turns out that the this is also the case for L\'evy Matrices.

\subsection{The exact equation for the mobility edge}

Considering the site $i_n$ of a given path, 
using the recursion relation~(\ref{eq:S}), one can rewrite the real part of the self-energy as:
\begin{displaymath}
S_{i_n} = \frac{h^2_{i_n,i_{n+1}}}{E - S_{i_n+1}} + \sum_{i_n^\prime = 1}^{N-2} \frac{h^2_{i_n,i_n^\prime}}{E - S_{i_n}^\prime} \, ,
\end{displaymath}
where the sum over $i_n^\prime$ runs over all the $N-2$ neighbors of $i_n$ except the sites $i_n$ and $i_{n+1}$ which belong to the path.
As a result, averaging over the quenched disorder the free 
energy~(\ref{eq:phi-EPAPS}) is written in terms of the  
largest eigenvalue, $\lambda(m,E)$, of the following transfer-matrix integral operator:
\begin{equation} \label{eq:Z-EPAPS}
\begin{split}
Z_{i_{n}} (S_{i_{n}}) &= (N-2) \int \textrm{d} S_{i_{n+1}} \, Z_{i_{n+1}} (S_{i_{n+1}}) \, \textrm{d} h_{i_n,i_{n+1}} \, P(h_{i_n,i_{n+1}}) \prod_{i_n^\prime}^{N-2} 
\left [ \textrm{d} h_{i_n,i_n^\prime} \, P(h_{i_n,i_n^\prime}) \, \textrm{d} S_{i_n^\prime}
L_{\mu/2}^{C(E),\beta(E)} (S_{i_n^\prime}) \right] \\
& \qquad \qquad \qquad \qquad \times \, 
\left| 
\frac{h_{i_n,i_{n+1}}}
{E - S_{i_{n+1}}}
\right|^{2m}
\, \delta \left( S_{i_n} - \frac{h^2_{i_n,i_{n+1}}}{E - S_{i_{n+1}}} - \sum_{i_n^\prime}^{N-2} \frac{h_{i_n^\prime}^2}{E - S_{i_n^\prime}} \right) \, ,
\end{split}
\end{equation}
where the factor $N-2$ accounts for the number of way one can choose the neighbors $i_{n+1}$ among all the $N-1$ neighbors of $i_n$ except the site $i_{n-1}$.
For large $R$ one has that $\phi(m,E) \simeq (1/m) \log \lambda(m,E)$. As a consequence, the mobility edge 
is found at the value of $E^\star$ where:
\begin{displaymath}
\left \{
\begin{aligned}
& \frac{1}{m} \log \lambda(m,E^\star) = 0 \, , \\ 
& \frac{\partial}{\partial m} \left[ \frac{1}{m} \log \lambda(m,E^\star) \right ] = 0 \, .
\end{aligned}
\right .
\end{displaymath}
This yields:
\begin{equation} \label{eq:ME-EPAPS}
\left \{
\begin{aligned}
& \lambda(m,E^\star) = 1 \, , \\ 
& \frac{\partial}{\partial m}  \lambda(m,E^\star) = 0 \, .
\end{aligned}
\right .
\end{equation}
Note that this is equivalent to study the stability of the localized phase
by checking whether a small imaginary part vanishes exponentially under iteration, as done in~\cite{abou-SM,JP-SM}
(in this case $\lambda(m,E)$ corresponds to the Lyapunov exponent of the imaginary part of the self-energy and
$m$ plays the role of the exponent of the power-law tails of its marginal probability distribution).
It is convenient to introduce the variable $S$ via the relation
\begin{displaymath}
\int \textrm{d} S \, \delta \left( S -  \sum_{i_n^\prime=1}^{N-2} \frac{h^2_{i_n,i_n^\prime}}
{E - S_{i_n^\prime}} \right)  = 1 \, .
\end{displaymath}
In the thermodynamic limit, $N-2 \simeq N-1 \simeq N$, 
according to the generalized central limit theorem $S$ has the same stationary distribution of the real part of the
self-energy, $L_{\mu/2}^{C(E),\beta(E)} (S)$:
\begin{displaymath}
\int \prod_{i_n^\prime}^{N-2} 
\left [ \textrm{d} h_{i_n,i_n^\prime} \, P(h_{i_n,i_n^\prime}) \, \textrm{d} S_{i_n^\prime}
L_{\mu/2}^{C(E),\beta(E)} (S_{i_n^\prime}) \right] \delta \left( S -  \sum_{i_n^\prime=1}^{N-2} \frac{h^2_{i_n,i_n^\prime}}
{E - S_{i_n^\prime}} \right)  \underset{N \to \infty}{\simeq} L_{\mu/2}^{C(E),\beta(E)} (S) \, .
\end{displaymath}
Eq.~(\ref{eq:Z-EPAPS}) then becomes:
\begin{equation} \label{eq:EV-EPAPS}
Z (X) = N  \int \textrm{d} h \, P(h) \, \textrm{d} S \, L_{\mu/2}^{C(E),\beta(E)} (S) \, \textrm{d} X^\prime Z (X^\prime) \,
\delta \left( X - S - \frac{h^2}{E - X^\prime} \right) \left| \frac{h}{E - X^\prime} \right |^{2m} \, .
\end{equation}
In order to solve the eigenvalue problem, we take the Fourier transform of both sides of Eq.~(\ref{eq:EV-EPAPS})
and obtain:
\begin{equation} \label{eq:almost}
\hat{Z}(k) = N \hat{L}_{\mu/2}^{C(E),\beta(E)} (k) \int \de h \, \de X^\prime \, P(h)  
\left | \frac{h}{E - X^\prime} \right |^{2 m} Z(X^\prime) \, 
e^{- i k h^2 / (E - X^\prime)} \, ,
\end{equation}
where $\hat{L}_{\mu/2}^{C(E),\beta(E)} (k)$ is the Fourier transform of the L\'evy stable distribution:
\begin{equation} \label{eq:levy-EPAPS}
\hat{L}_{\mu/2}^{C(E),\beta(E)} (k) = \exp \left[ - C(E) |k|^{\mu/2} \left( 1 + i 
\beta(E) \tan \left( \frac{\pi \mu}{4} \right) \textrm{sign}(k) \right) \right] \, .
\end{equation}
Using the fact that:
\begin{equation} \label{gamma}
\int_0^{\infty} 
\de x \, \frac{e^{i k x}}{x^a} = e^{i \frac{\pi}{2} (1 - a) \textrm{sign}(k)} |k|^{a-1} \Gamma(1-a) \, ,
\end{equation}
we can easily integrate over the disorder matrix element distribution obtaining:
\begin{displaymath}
\int_0^{\infty} \de h \, P(h)  
\, | h|^{2m} \, e^{- i k h^2 / (E - X^\prime)} = \frac{\mu}{2N} \, \Gamma( m - \mu/2) \, 
e^{- i \frac{\pi}{2} (m - \mu/2) \textrm{sign}
(k (E - X^\prime))} \left | \frac{k}{E - X^\prime} \right |^{\mu/2 - m} \, .
\end{displaymath}
Plugging the result above into Eq.~(\ref{eq:almost}) we  get:
\begin{equation} \label{eq:final}
\hat{Z}(k) = \frac{\mu}{2} \, \Gamma( m - \mu/2) \, |k|^{\mu/2 - m} \, 
\hat{L}_{\mu/2}^{C(E),\beta(E)} (k) 
\int_{-\infty}^{+\infty} \de X^\prime \,
\frac{Z(X^\prime)}{|E - X^\prime|^{m + \mu/2}} \, e^{- i \frac{\pi}{2} (m - \mu/2) \textrm{sign}
(k (E - X^\prime))} \, .
\end{equation}
Note that this is exactly the same equation found in~\cite{JP-SM}, where the authors studied the stability of the localized phase
by determining the Lyapunov exponent of the imaginary part of the self-energy under iteration.

It is convenient to replace $Z(X^\prime)$ by the inverse Fourier transform of $\hat{Z}(k^\prime)$.
We then perform the integral over $\de X^\prime$ by separating it into two pieces, 
and changing variable $E-S \to z$ in the interval $(-\infty,E)$ and $S-E \to z$ in the interval 
$(E,+\infty)$. Using again Eq.~(\ref{gamma}) we find:
\begin{displaymath}
\begin{split}
&\int_{-\infty}^{+\infty} \de X^\prime \, \frac{e^{i k^\prime X^\prime}}{|E - X^\prime|^{m + \mu/2}} 
e^{- i \frac{\pi}{2} (m - \mu/2) \textrm{sign} (k (E - X^\prime))} = \\
&e^{i k^\prime E} \, \Gamma(1 - m - \mu/2) \, |k^\prime|^{m + \mu/2 - 1} \left[ 
e^{- i \frac{\pi}{2} (m - \mu/2) \textrm{sign} (k)} \, e^{-i \frac{\pi}{2}
( 1 - m - \mu/2) \textrm{sign} (k^\prime)} + e^{i \frac{\pi}{2} (m - \mu/2) \textrm{sign} (k)}
\, e^{i \frac{\pi}{2} ( 1 - m - \mu/2) \textrm{sign} (k^\prime)} \right] \, .
\end{split}
\end{displaymath}
We plug back this last result into Eq.~(\ref{eq:final}) and get:
\begin{displaymath}
\begin{split}
\hat{Z}_+ (k) &= \frac{\mu}{2} \, \Gamma( m - \mu/2) \, \Gamma(1 - m - \mu/2) \, |k|^{\mu/2 - m} \, 
\hat{L}_{\mu/2}^{C(E),\beta(E)} (k) \left [ \sin \left( \frac{\pi \mu}{2} \right) I_+
+ \sin \left( \pi m \right) I_- \right ] \, ,\\
\hat{Z}_- (k) &= \frac{\mu}{2} \, \Gamma( m - \mu/2) \, \Gamma(1 - m - \mu/2) \, |k|^{\mu/2 - m} \,
\hat{L}_{\mu/2}^{C(E),\beta(E)} (k) \left [ 
\sin \left( \pi m \right) I_+ + 
\sin \left( \frac{\pi \mu}{2} \right) I_- \right] \, ,
\end{split}
\end{displaymath}
where $I_+$ and $I_-$ are defined as:
\begin{displaymath}
\begin{split}
I_+ &= \int_{0}^{+\infty} \frac{\de k^\prime}{\pi}
\, e^{i k^\prime E} \, |k^\prime|^{m + \mu/2 - 1} \, \hat{Z}(k^\prime) \, , \\
I_- & = \int_{-\infty}^0 \frac{\de k^\prime}{\pi}
\, e^{i k^\prime E} \, |k^\prime|^{m + \mu/2 - 1} \, \hat{Z}(k^\prime) \, .
\end{split}
\end{displaymath} 
We also have defined $\hat{Z}_+ (k)$ and $\hat{Z}_- (k)$ as the function $\hat{Z} (k)$
restricted to the regions $k>0$ and $k<0$ respectively, and introduce the
coefficients:
\begin{displaymath}
\begin{split}
K_{m,\mu} &= \frac{\mu}{2} \, \Gamma \left( m - \frac{\mu}{2} \right) \, \Gamma \left(1 - m - \frac{\mu}{2} \right) \, , \\
s_\mu & = \sin \left( \frac{\pi \mu}{2} \right) \,  , \\
s_m & = \sin \left( \pi m \right) \, ,
\end{split}
\end{displaymath}
as done in the main text. 
We multiply both $\hat{Z}_+ (k)$ and $\hat{Z}_- (k)$ by  $e^{i k E} \, 
|k|^{m + \mu/2 - 1}$ and integrate them over $\de k / \pi$ in the intervals $(0,+\infty)$ and $(-\infty,0)$ respectively. 
We then obtain:
\begin{equation} \label{eq:2b2-EPAPS}
\begin{split}
I_+ &=  K_{m,\mu} \left ( s_\mu I_+ + s_m I_- \right ) \ell_+  \, , \\
I_- & =  K_{m,\mu} \left ( s_m I_+ + s_\mu I_- \right) \ell_- \, ,
\end{split}
\end{equation}
where $\ell_+$ and $\ell_-$ are defined as:
\begin{displaymath}
\begin{split}
\ell_+ & = \int_{0}^{+\infty} \frac{\de k}{\pi}
\, e^{i k E} \, |k|^{\mu - 1} \hat{L}_{\mu/2}^{C(E),\beta(E)} (k) \, , \\
\ell_- & = \int_{-\infty}^0 \frac{\de k}{\pi}
\, e^{i k E} \, |k|^{\mu - 1} \hat{L}_{\mu/2}^{C(E),\beta(E)} (k) \, .
\end{split}
\end{displaymath}
The $2 \times 2$ linear system~(\ref{eq:2b2-EPAPS}) 
only has non-trivial solutions different from zero if the determinant of the
matrix of the coefficients vanishes.
Hence the equation~(\ref{eq:ME-EPAPS}) for the mobility edge reads:
\begin{equation} \label{determinant}
K_{m,\mu}^2 \ell_+ \ell_- \left[ s_\mu^2 - s_m^2 \right] 
- K_{m,\mu} \left(\ell_+ + \ell_- \right) s_\mu + 1 = 0 \, .
\end{equation}
Using the specific form of $\hat{L}_{\mu/2}^{C(E),\beta(E)} (k)$, Eq.~(\ref{eq:levy-EPAPS}), it is straightforward
to show that $\ell_- = \ell_+^*$. 
Interestingly enough, we remark that, as for the Anderson model on the Bethe lattice~\cite{abou-SM}, 
the l.h.s. of Eq.~(\ref{determinant}) is a symmetric function
of $m$ around $m=1/2$. This implies that,  
if a solution of Eq.~(\ref{determinant}) exists, 
then the stationary condition $\partial \lambda(m,E^\star) / \partial m = 0$ can 
only be verified for $m=1/2$.
Hence, Eq.~(\ref{determinant}) finally becomes Eq.~(2) of the
main text.

We have solved Eq. (\ref{determinant})---together with Eqs.~(\ref{eq:C-self-EPAPS}) and Eqs.~(\ref{eq:beta-self-EPAPS})---numerically 
for several values of $\mu \in (0,1)$ and $E$ (and for $m=1/2$), and obtained the phase 
diagram of fig.~1 of the main text.

\section{Super-symmetric formalism}
We give here more details on the super-symmetric formalism we discussed in the main text. 
The super-symmetric method to study random matrices is well established by now~\cite{efetov-SM}. Moreover, 
our derivations are along the lines of  the ones developed by Mirlin and Fyodorov for the connectivity matrix and the Anderson localization of 
finite connectivity random graphs~\cite{mirlinbethe-SM}. For completeness, we present the main ideas and technical steps. 
A full derivation will be presented elsewhere~\cite{long-SM}. 
\subsection{Action in terms of $\rho(\Phi)$}
As stated in the main text, the starting point is the field theory $Z=\int \prod_i \textrm{d}\Phi_i e^{S[\Phi_i]}$, with the action~\cite{efetov-SM}:
\[
S=\frac i 2 \left(\sum_{l,m} \Phi_l^\dagger \mathcal{L} (E\delta_{lm}-h_{lm})\Phi_m +\sum_l \Phi_l^\dagger \Phi_l \, \frac{r+i0^+}{N} \right) \, .
\] 
The field $\Phi_i$ is a eight-component super-vector $(\Phi^{(1)}_i,\Phi^{(2)}_i)=
(S_i^{a},S_i^{b},\chi_i,\chi_i^*,P_i^{a},P_i^{b},\eta_i,\eta_i^*)$, where each of the four component super-vector $\Phi^{(1,2)}_i$
is formed by two real and two Grassman variables. The matrix $\mathcal{L}$ is diagonal with elements $(1,1,1,1,-1,-1,-1,-1)$. 
In order to average over the distribution of the matrix elements one has to compute:
\[
\prod_{l<m}\overline{ e^ {-i  h_{lm}\Phi_l^\dagger \mathcal{L} \Phi_m}} \prod_{l}\overline{e^ {-\frac i 2  h_{ll}\Phi_l^\dagger \mathcal{L} \Phi_l} \, .
}
\]
Note that we have used explicitly that $\Phi_l^\dagger \mathcal{L} \Phi_m=\Phi_m^\dagger \mathcal{L} \Phi_l$. 
Using that 
\[
\overline{e^ {-i  h_{lm}\Phi_l^\dagger \mathcal{L} \Phi_m}}\simeq 1+\frac \mu N \int \frac{\textrm{d}h_{lm}}{2|h_{lm}|^{1+\mu}} 
\left(e^ {-i  h_{lm}\Phi_l^\dagger \mathcal{L}\Phi_m}-1\right) \, .
\]
one finds an action 
which reads (up to subleasing terms in $N$, that do not play any role for our derivations):
\[
S_{a}=\frac i 2 \left(\sum_{l} E \Phi_l^\dagger \mathcal{L} \Phi_l +\sum_l \Phi_l^\dagger \Phi_l \, \frac{r+i0^+}{N} \right)
+\frac{\mu}{2N}\sum_{lm} \int \frac{\textrm{d}h}{2|h|^{1+\mu}} \left(e^ {-i  h\Phi_l^\dagger \mathcal{L}\Phi_m}-1\right) \, .
\] 
Introducing the function $\rho(\Phi)=\frac 1 N \sum_i \delta(\Phi-\Phi_i)$ one can rewrite $S_a$ as
\begin{displaymath}
S_a=\frac i 2 NE \!  \int \! \textrm{d} \Phi \rho(\Phi)   \Phi^\dagger \mathcal{L} \Phi + \frac i 2 (r+i0^+) \! \int \! \textrm{d}\Phi   \rho(\Phi) \Phi^\dagger \Phi
+\frac i 2N \!\int\! \textrm{d} \Phi \textrm{d}\Psi  \rho(\Phi) C(\Phi^\dagger \mathcal{L} 
\Psi)\rho(\Psi) \, ,
\end{displaymath}
the function $C(y)$ being the one defined in the main text:
$C(y)=\mu \int \frac{\textrm{d}x}{ 2 |x|^{1+\mu}}[\exp(-ixy)-1]$.
It still remains to integrate over all $\Phi_i$s.
Since the action depends on the $\Phi_i$s through $\rho(\Phi)$ only, one can first integrate over 
all $\Phi_i$s that correspond to the same $\rho(\Phi)$. This leads to an additional entropic-like term in the 
action $-N \!\int\! \textrm{d}\Phi  \rho(\Phi)\log  \rho(\Phi)$ (the computation is standard even though generically is not done with super-field).   
The final result is that the field theory has been now transformed in a new one:
$Z=\int {\mathcal D} \rho(\Phi) e^{S[ \rho]}$ where the integral is over all normalized $ \rho(\Phi)$ with the action reading
\begin{equation} \label{eq:SrhoSS}
S[\rho] =\frac i 2 NE \!  \int \! \textrm{d} \Phi \rho(\Phi)   \Phi^\dagger \mathcal{L} \Phi + \frac i 2 (r+i0^+) \! \int \! \textrm{d}\Phi   \rho(\Phi) \Phi^\dagger \Phi
- N \!\int\! \textrm{d}\Phi  \rho(\Phi)\log  \rho(\Phi)+\frac N 2 \!\int\! \textrm{d} \Phi \textrm{d}\Psi  \rho(\Phi) C(\Phi^\dagger \mathcal{L} 
\Psi)\rho(\Psi)\, .
\end{equation} 
The interest of this formulation is that because of the $N$ which can be factored out in the action, one can 
evaluate it by the saddle-point method. 

\subsection{Relationship between $\rho(\Phi)$ and $R(\Sigma)$}
Before discussing the corresponding equation it is useful to 
remark that the average value of $\rho(\Phi)$, which corresponds to the saddle point of the previous integral, has a particularly illuminating expression in terms of the 
distribution of the local self-energy $R(\Sigma)$~\cite{mirlinsigma-SM}. Before that the average over the disorder is performed, the field theory is Gaussian.
Hence, by integrating all fields but $\Phi_i$ one remains with a Gaussian integral to handle.
Using that the field theory is constructed in such a way that $\langle \Phi_i^{(1)\dagger} \Phi_j^{(1)} \rangle=4iG_{ij}$ 
and $\langle \Phi_i^{(2)\dagger} \Phi_j^{(2)} \rangle=4iG_{ij}^*$ ($G_{ij}$ is the resolvent 
and $\langle \cdot \rangle$ denotes the average over the field theory at fixed disorder) the average $\langle \delta(\Phi-\Phi_i)\rangle $
turns out to be 
the Gaussian measure on $\Phi_i$. By collecting all terms and averaging over the disorder one finds:
\[
\overline{\langle \rho(\Phi) \rangle}=\frac 1 N\sum_i \overline{\exp\left(\frac i 2 \Phi^\dagger \mathcal{L} \Phi (E-\Re \Sigma_{ii})+\frac 1 2 \Phi^\dagger \Phi \Im \Sigma_{ii}
\right) \, .
} 
\] 
By introducing the distribution of the local self-energy $R(\Sigma)$, one gets the expression quoted in the text and already derived in~\cite{mirlinsigma-SM}:
\begin{equation}\label{expression}
\overline{\langle \rho(\Phi) \rangle}=\int \textrm{d} \Sigma R(\Sigma) \exp\left(\frac i 2 \Phi^\dagger \mathcal{L} \Phi (E-\Re \Sigma)+\frac 1 2 \Phi^\dagger \Phi \Im \Sigma
\right) \, .
\end{equation}
Since the field theory can be solved by the saddle point method, 
the saddle-point value of $\rho(\Phi)$ has to satisfy the previous equation as stated in the text. We show below that this is indeed the case. 

\subsection{Equation on $R(\Sigma)$}
As discussed in the main text, as well as in the first section of the SM, 
it was shown in~\cite{JP-SM, burda-SM,metz-SM,benarous-SM} that the probability distribution of the local self energy, $R(\Sigma)$ 
[or equivalently of the diagonal element of the resolvent, Eq.~(1) of the main text] is obtained in the large-$N$ limit from the equation:
\begin{equation}
\label{eq:S}
\Sigma_{ii} \overset{\rm d}{=} 
\sum_{j=1}^N h_{ij}^2 G_{jj}
= \sum_{j=1}^N \frac{h_{ij}^2}{E -\Sigma_{jj}} \, , 
\end{equation}
where $\overset{\rm d}{=}$ denotes the equality in distribution between random variables and $E$ contains an infinitesimally small imaginary part $i0^+$.
All correlations between the terms on the RHS of the previous relation can be neglected in the thermodynamic limit.
This leads to a self-consistent equation on $R(\Sigma)$. 
For a full explanation of its derivation, see~\cite{benarous-SM}. Here 
we just sketch how one can obtain an identity on its generating function. 

The first idea is that since the correlations between the terms on the RHS can be neglected, as it was shown in~\cite{benarous-SM}, $\Sigma_{ii}$ is a sum of a large number or 
heavy tailed i.i.d.~variables and, hence, it's a complex L\'evy random variable. We consider its generating function:
\[\int \! \textrm{d} \Sigma R(\Sigma) e^{-iX_1\Sigma+iX_2\Sigma^*}=\prod_{j}\overline{e^{-ih_{ij}^2 (X_1G_{ij}-X_2G_{ij}^*)}} \, ,
\]
where $X_1$ and $X_2$ are two real variables. 
The RHS can be computed in the following way:
\[
\prod_{j}\overline{e^{-ih_{ij}^2 (X_1G_{ij}-X_2G_{ij}^*)}}=\left(1+\frac \mu N \int \! \textrm{d} \Sigma R(\Sigma) \! \int \! \frac{\textrm{d} h}{2|h|^{1+\mu}}
\left[ e^{-ih^2 \left( \frac{X_1}{E -\Sigma}-\frac{X_2}{E -\Sigma^*}\right)}
-1\right]\right)^N \, .
\]
Henceforth we neglect all the subleading (vanishing) terms in $1/N$. This allows one to derive the identity:
\begin{equation}\label{eqR}\int \! \textrm{d} \Sigma R(\Sigma) \, e^{-iX_1\Sigma+iX_2\Sigma^*}
=\exp\left(\mu \! \int \!  \textrm{d} \Sigma R(\Sigma) \! \int \! \frac{\textrm{d}h}{2|h|^{1+\mu}}
\left[ e^{-ih^2 \left(\frac{X_1}{E -\Sigma}-\frac{X_2}{E -\Sigma^*}\right)}
-1\right]\right) \, .
\end{equation}
This is an implicit version of the self-consistent equation satisfied by $R(\Sigma)$, which also defines $R(\Sigma)$ as a complex L\'evy stable distribution~\cite{benarous-SM}.

We show in the following that this same result also follows directly from the saddle point equation~(\ref{expression}) on $\rho(\Phi)$. 
By extremizing the action~(\ref{eq:SrhoSS}) on $\rho(\Phi)$ and taking into account the normalization condition on $\rho(\Phi)$, at leading order in $N$ one finds:
\begin{displaymath}
\rho(\Phi)  =\exp\left(\frac i 2 E\Phi^\dagger \mathcal{L} \Phi+ \!\int\!  \textrm{d}\Psi  C(\Phi^\dagger \mathcal{L} 
\Psi)\rho(\Psi) \right) \, .
\end{displaymath}
By plugging the expression~(\ref{expression}) into the previous equation, one can perform the integral over $\Psi$:
\begin{displaymath}
\begin{split}
& \int \textrm{d} \! \Sigma R(\Sigma) \exp\left(\frac i 2 \Phi^{(1)\dagger} \Phi^{(1)} (E-\Sigma)-\frac i 2 \Phi^{(2)\dagger} \Phi^{(2)} (E-\Sigma^*)
\right) = \\
& \qquad \qquad \exp\left(\frac i 2 E \left( \Phi^{(1)\dagger} \Phi^{(1)}-\Phi^{(2)\dagger} \Phi^{(2)} \right) + 
\mu \! \int \! \textrm{d} \Sigma R(\Sigma)
 \!\int\!  \frac{\textrm{d}h}{2|h|^{1+\mu}} 
\left[ e^{-ih^2 \left(\frac{\Phi^{(1)\dagger} \Phi^{(1)}}{E -\Sigma}-\frac{\Phi^{(2)\dagger} \Phi^{(2)}}{E -\Sigma^*}\right)}
-1\right]\right) \, .
\end{split}
\end{displaymath}
This expression has to be valid for any $\Phi^{(1)\dagger} \Phi^{(1)}$ and $\Phi^{(2)\dagger} \Phi^{(2)}$, hence it 
defines a self-consistent on $R(\Sigma)$ which actually coincides with Eq.~(\ref{eqR}) established previously. 
This result shows that our super-symmetric formalism is in agreement with previous exact results:  
$R(\Sigma)$ is the complex L\'evy stable distribution obtained rigorously in~\cite{benarous-SM}. 

\section{Dyson Brownian Motion and the regime $1<\mu<2$}
As we explained in the main text, we can show that the level statistics for $1<\mu<2$ is the one
of GOE matrices under the hypothesis that all Wigner matrices with the same  
heavy tails are characterized by the same level statistics (a very reasonable assumption). 

Our strategy consists first in modifying the distribution $P(h_{ij})$ of the matrix elements into $(1-\epsilon) P(h_{ij})+\epsilon N^{1/\mu}W(N^{1/\mu}h_{ij})$
where $W(x)$ is a Gaussian distribution with unit variance.  This does not alter the fat tails of the matrix elements
and allows focusing on random matrices $\mathcal{H}_{\epsilon}=(1-\epsilon) \mathcal{H}+\epsilon \mathcal{W}$
where $\mathcal{H}$ is a LM and $\mathcal{W}$ a very small GOE matrix
whose elements have exactly the same scaling with $N$ than the ones of $\mathcal{H}$. 
The level statistics of $\mathcal{H}_{\epsilon}$---and, by the previous assumption, of $\mathcal{H}$---can 
be obtained using the Dyson Brownian Motion. 
Since this is a very well known technique, we refer to the literature for an introduction~\cite{metha-SM,yau-SM} and just reproduce the main steps and ideas needed for our argument.   

Let's denote $\lambda_i(t)$ the eigenvalues of the matrix $\mathcal{H}_{t}=(1-t) \mathcal{H}+t\mathcal{W}$.
For $t=0$ these coincides with the eigenvalues of $\mathcal{H}$ and for $t=\epsilon$ with the ones of $\mathcal{H}_{\epsilon}$. 
The idea behind the Dyson Brownian Motion technique is to transform the interpolation $t=0\rightarrow \epsilon$ in a stochastic process
on the eigenvalues. The main trick is the property that a Gaussian variable can be considered as the sum of two independent Gaussian variables. 
This allows one to describe the change from the $\mathcal{H}_{t}$ ensemble to the $\mathcal{H}_{t+\rm{d} t}$ ensemble as an addition of a GOE matrix 
(with a suitable variance) and a rescaling. Taking an infinitesimal $\rm{d}t$ allows one to use perturbation theory: the final result that goes under 
the name of Dyson Brownian Motion is a stochastic evolution equation on the eigenvalues. The eigenvalues of $\mathcal{H}$ are simply the initial conditions for this 
process, whereas the ones 
of  $\mathcal{H}_{\epsilon}$ are the values obtained after a ``time'' $t=\epsilon$. 

The DBM admits a stationary distribution for the $\lambda_i(t)$s which is simply the GOE distribution. The crucial question is then 
whether the $\lambda_i(t)$ have enough ``time'' to equilibrate to their equilibrium (GOE) probability measure.   
What was conjectured already by Dyson and proved in great generality in recent years~\cite{yau-SM} is that the relaxation timescale to obtain local equilibration in 
the bulk of the spectrum to GOE statistics scales as $N^{1/\mu}/N$ (with the scaling we have considered in this work).

In consequence, for any $\mu \in (1,2)$, for any finite value of $t$, in particular $t=\epsilon$, the statistics of the  
eigenvalues in the bulk of the spectrum converges to the GOE one in the large-$N$ limit. Using the assumption that all matrices with the same  
heavy tails are characterized by the same level statistics, we then find GOE level statistics for all matrices $\mathcal{H}_{\epsilon}$,
in particular $\mathcal{H}=\mathcal{H}_{\epsilon=0}$. 

\section{Numerical results for $\mu \in (1,2)$}

In this section we provide several numerical results obtained from exact diagonalizations
of L\'evy Matrices in the range $\mu \in (1,2)$, for several system sizes 
$N = 2^m$, with $m$ from $8$ to $14$. 
As explained in the main text, the data are averaged over (at least) $2^{22-m}$
realization of the disorder. The energy spectrum is resolved in $64$ small
intervals $\nu$, centered around the energies $E_\nu = \langle \lambda_n \rangle_\nu$. 

\begin{figure}
\includegraphics[width=0.42\textwidth]{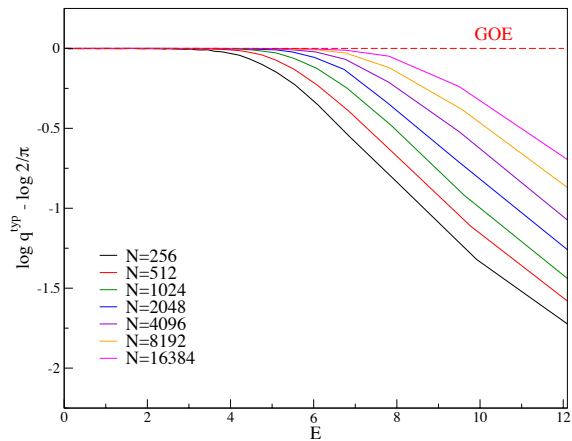}%
\caption{\label{fig:q-1.5}
$\ln(q^{typ}/q^{typ}_{GOE})$ as a function of the energy $E$ for different system sizes for $\mu = 1.5$.}
\end{figure}

In fig.~\ref{fig:q-1.5} we plot $q^{typ}_\nu$ as a function of $E_\nu$ for
$\mu = 1.5$ and for different system sizes,
averaged over samples and eigenstates within each energy window.
Although at high energies our numerical data are still quite far from full convergence,
it is clear that $q^{typ}_\nu$
evolves towards the GOE universal value, $q^{typ}_{GOE} = 2/\pi$ 
in all energy windows as $N$ is increased.

\begin{figure}
\includegraphics[width=0.5\textwidth]{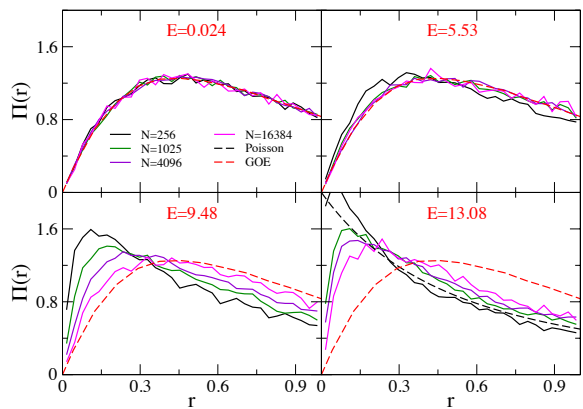}%
\caption{\label{fig:Pr-1.5} Probability distribution of the gap ratio for $\mu = 1.5$,
for different system sizes, and for four different values of the energy. The Poisson
and GOE counterparts of $\Pi(r)$ are also shown.
Top-left panel: $E = 0.024$; The entire probability distribution  is described by the
GOE. Top-right panel: $E = 5.53$; $\Pi(r)$ converges to the GOE distribution for $N$ large enough.
Bottom-left panel: $E = 9.48$; $\Pi (r)$ evolves towards the GOE distribution
as $N$ is increased, although we are not able to observe full convergence.
Bottom-right panel: $E = 13.08$; $\Pi (r)$ is very far from convergence even for the largest system
size considered, although it slowly evolves towards the GOE distribution.}
\end{figure}

In fig.~\ref{fig:Pr-1.5} we show the probability distribution of the gap ratio, $\Pi(r)$,
for $\mu=1.5$, for different system sizes, and for four different values of the energy.
In the case of Poisson statistics the probability distribution of the gap ratio is given by
$\Pi (r) = 2/(1+r)^2$, while the counterpart of $\Pi (r)$ 
corresponding to GOE statistics has been computed exactly in~\cite{PR-GOE}.
Level repulsion in the GOE spectra manifests itself in the vanishing of the probability
distribution at $r=0$.
As expected, at small enough energy 
the entire probability distribution $\Pi(r)$ is described by GOE. 
Finite size effects are stronger at higher energies.
In particular for $E \simeq 13.08$ (bottom-right panel of \ref{fig:Pr-1.5}), 
$\Pi(r)$ is still quite far from convergence 
even at the largest system size. Nevertheless it is clear that the distribution
of the gap ratio is slowly
evolving towards the GOE distribution as $N$ is increased. 

\begin{figure}
\includegraphics[width=0.42\textwidth]{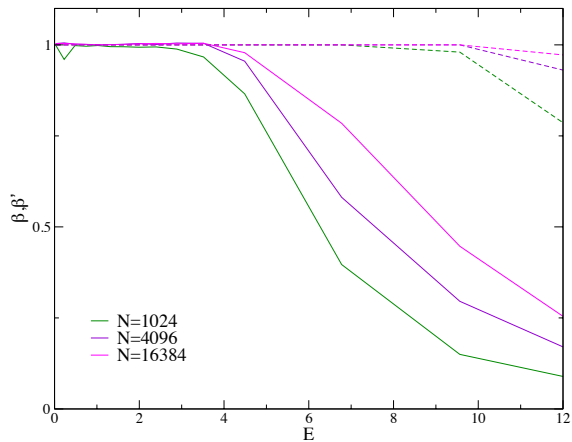}%
\caption{\label{fig:IPR-1.5} Exponents $\beta$ (continuous lines) and $\beta^\prime$ (dashed lines) describing the 
scaling with $N$ of the typical value of the Inverse Participation Ratio and of the
support set as a function of the energy for $\mu = 1.5$.}
\end{figure}

Numerical results on the Inverse Participation Ratios (IPR) are
coherent with previous findings. 
The IPR of the eigen-function ${n}$ is defined
as: $\Upsilon_{2,n} = \sum_{i=1}^N |\langle{i}|{n}\rangle |^4$.
In fig.~\ref{fig:IPR-1.5} we plot the energy dependence of the 
exponent $\beta= \langle \ln \Upsilon_{2,n} \rangle_\nu/\ln N$ describing the scaling of the typical value
of the IPR with the system size.
At small enough energies we find $\beta \simeq 1$, corresponding
to the standard scaling of the IPR for fully delocalized states.
As mentioned above, finite size effects are stronger at higher energies.
We indeed observe that $\beta$ 
decreases as the energy grows for a fixed system size $N$.
Nevertheless, at fixed energy, $\beta$ increases as the system size is
increased and seems to approach the standard value $1$ in all
energy windows. 
This is confirmed by the numerical results  on the 
the support set, recently introduced
in~\cite{scardicchiosupport-SM} as a tough measure wave-functions ergodicity.
For an eigenvector ${n}$ with sites ordered 
according to $|\langle{i}|{n}\rangle| > |\langle{i+1}|{n}\rangle|$, it is defined as the sets of sites $i < S_\epsilon^{(n)}$ such that
$\sum_{i=1}^{S_\epsilon^{(n)}} |\langle{i}|{n}\rangle|^2 \le 1 - \epsilon < \sum_{i=1}^{S_\epsilon^{(n)}+1} |\langle{i}|{n}\rangle|^2$.
The scaling of $\langle S_\epsilon^{(n)} \rangle$ for $N \to \infty$ and $\epsilon$ arbitrary small but finite allows to
discriminate between the extended and the localized regimes, as $S_\epsilon^{(n)}$ is $N$-independent for localized wave-functions while
it diverges for $N \to \infty$ for delocalized states.
The exponent
$\beta^\prime = \ln \langle S_\epsilon^{(n)} \rangle_\nu / \ln N$, describing the scaling of the support set
at large $N$ is also shown in 
fig.~\ref{fig:IPR-1.5}. Its behavior is very similar to the one of $\beta$ described above.
However the support set is apparently a sharper measure of wave-functions ergodicity compared to the IPR, 
as the values of  $\beta^\prime$
are much closer to 
$1$ in all energy windows.

Similar results are obtained for $\mu = 1.1$, confirming that 
for $\mu \in (1, 2)$ all eigenstates of LMs are extended and the
level statistics is described by GOE in the whole spectrum.
Nevertheless finite size effect become stronger as $\mu$ is lowered and 
can be extremely important at high energies, where one needs to consider relatively large
$N$ to observe full converges towards GOE. 

\section{Numerical results for $\mu \in (0,1)$}

According to the expression~(\ref{eq:Ph}) of the probability distribution of the entries of our LMs, 
each row (or column) of $\mathcal{H}$ has $O(N)$ elements of $O(N^{-1/\mu})$ and $O(1)$ elements of $O(1)$, 
ensuring a well-defined thermodynamic limit.
However, 
the largest element of the whole matrix (which contains $N^2$ terms)
is of order $N^{1/\mu}$.
As a consequence, 
the range of variability of the matrix elements
goes %
from $O(N^{-1/\mu})$ to $O(N^{1/\mu})$, 
which is, for large enough system sizes and for $\mu<1$, extremely broad.
This could affect the numerical precision of our results.
In order to overcome this issue, we have introduced a
cut-off on large matrix elements scaling as $\Lambda N^{1/\mu}$, where $\Lambda$ is
a constant much larger than $1$.
Since we are only interested in the properties of LMs for 
energies of $O(1)$, the presence of such cut-off does not have
any influence on our numerical results (provided that $\Lambda$ is large enough).

Furthermore, 
in order to exploit the ``sparse-like'' character of LMs, we introduce a cut-off $\gamma$ (very small but finite) on small matrix elements which allows 
to transform $\mathcal{H}$ in a sparse Erd\"os-R\'enyi random matrix constituted by the
backbone of large entries~\cite{metz-SM}.
This allows to simplify and speed-up the numerical calculations, since numerical routines for exact diagonalization
are faster for sparse matrices.
The probability distributions of the entry thus becomes:
\begin{displaymath}
P_N^{(\gamma,\Lambda)} (h_{ij}) = p_N^{(\gamma)} \delta(h_{ij}) 
+ \left( 1 - p_N^{(\gamma)} \right) \theta \left( \gamma < |h_{ij}| < N^{1/\mu} \Lambda \right) \frac{C_N^{(\gamma,\Lambda)}}{2 |h_{ij}|^{1+\mu}} \, ,
\end{displaymath}
where $p_N^{(\gamma)} = 1 - 1/(N \gamma^\mu)$ and $C_N^{(\gamma,\Lambda)} = \mu/[\gamma^{-\mu} - N^{-1} \Lambda^{-\mu}]$.

We have performed exact diagonalizations of such random matrices for several system sizes 
$N = 2^m$, with $m$ from $8$ to $15$, and for $\Lambda = 2^{15}$.
Data are averaged over (at least) $2^{22-m}$
realization of the disorder. The energy spectrum is resolved in $64$ small
intervals $\nu$, centered around the energies $E_\nu = \langle \lambda_n \rangle_\nu$.
In order to make sure that the cut-off on small entries is small enough to reproduce the $\gamma \to 0$ limit, 
we have considered different values of $\gamma$
($\gamma = 10^{-3}$, $10^{-4}$, and $5 \cdot 10^{-5}$) 
and checked that the data become independent of it (within our numerical accuracy).

In the following we complement the discussion of the main text
with more details and results obtained from exact diagonalizations 
for $\mu \in (0,1)$.
We will focus more specifically on $\mu=0.5$. 
Similar results are found for $\mu = 0.8$ and $\mu = 0.3$, although
finite size effects becomes bigger as $\mu$ is decreased and the
crossover region gets broader. 

\begin{figure}
\includegraphics[width=0.42\textwidth]{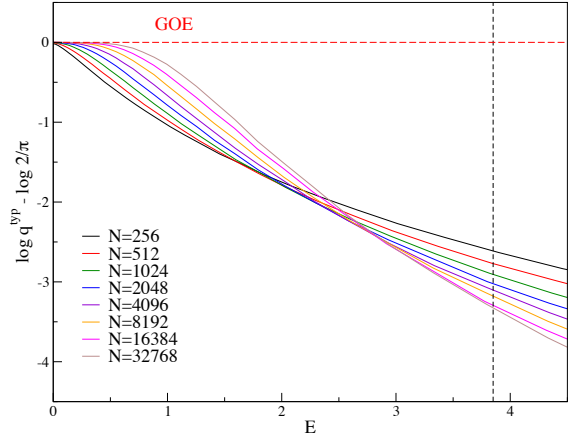}%
\caption{\label{fig:q-0.5}
$\ln(q^{typ}/q^{typ}_{GOE})$ as a function of the energy $E$ for different system sizes for $\mu = 0.5$.}
\end{figure}

In fig.~\ref{fig:q-0.5} we plot $q^{typ}_\nu$ as a function of $E_\nu$ for
$\mu = 0.5$ and for different system sizes,
averaged over samples and eigenstates within each energy window.
For small (resp. large) energies we recover, as expected, the universal values
$q_P^{typ} = 2/\pi$
(resp. $q_P^{typ} \to 0$) corresponding to GOE (resp. Poisson) statistics.
As mentioned in the main text, the curves corresponding to different values of $N$ 
seem to cross much before the localization transition, which can be computed analytically
and should occur at $E^\star \simeq 3.85$. 
However, our numerical data on $q^{typ}$ are extremely clean and allow to 
observe that the crossing point is actually slowly drifting
towards higher values of the energy (and most probably converging to $E^\star$ in the thermodynamic limit).

\begin{figure}
\includegraphics[width=0.5\textwidth]{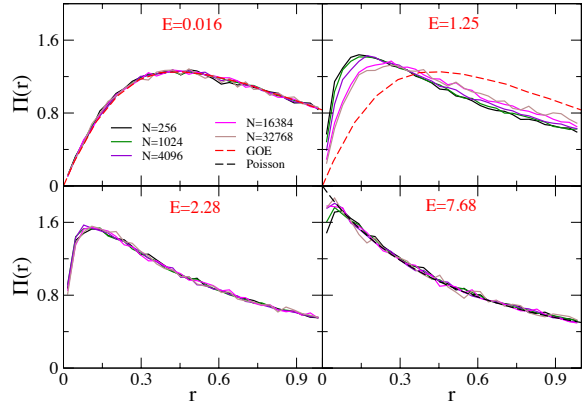}%
\caption{\label{fig:Pr-0.5} Probability distribution of the gap ratio for $\mu = 0.5$, 
for different system sizes, and for four different values of the energy. The Poisson
and GOE counterparts of $\Pi(r)$ are also shown.
Top-left panel: $E = 0.016$; The entire probability distribution  is described by the 
GOE. Top-right panel: $E = 1.25$; $\Pi(r)$ evolves towards the GOE distribution 
as $N$ is increased, although we are not able to observe full convergence.
Bottom-left panel: $E = 2.28$; $\Pi (r)$ seems to be described by a $N$-independent 
non-universal distribution. 
Bottom-right panel: $E = 7.68$; $\Pi (r)$ converges to the Poisson distribution
for large $N$.}
\end{figure}

In fig.~\ref{fig:Pr-0.5},
we show the probability distribution of the gap ratio, $\Pi(r)$,
for $\mu=0.5$, for different system sizes, and for four different values of the energy.
As expected, for small enough energies (e.g., $E \simeq 0.016$, top-left panel) the entire probability distribution is described
by GOE statistics. Conversely, for high enough energies (e.g., $E \simeq 7.68$, bottom-right panel), in the localized regime,
the data nicely approach the Poisson distribution $\Pi(r) = 2/(1+r)^2$---except for very small values of $r$ where
convergence is exponentially slow due to finite size effects.
For moderately high energies (e.g., $E=1.25$, top-right panel), $\Pi (r)$ evolves towards the GOE distribution
as $N$ is increased, although we are not able to observe full convergence for the largest system size.
Finally, for energies in the crossover region (e.g., $E=2.28$, bottom-left panel), 
one seems to observe that $\Pi (r)$ is described by a stationary (i.e., $N$-independent)
and non-universal---neither GOE nor Poisson---distribution, as observed in~\cite{JP-SM}. 
Nevertheless, if one analyzes carefully the numerical data, focusing, for instance, 
on the behavior of $\Pi(r)$ at small $r$, one realizes that 
$\Pi(r)$ evolves in a non-monotonic way: for system sizes smaller than the crossover size, $N<N_m \simeq 1200$ (see
the inset of fig.~3 of the main text), it evolves 
towards the Poisson distribution,
while for large system sizes, $N>N_m$, it commences to approach the GOE distribution.
However, it is evident that if one ignored the existence of the crossover scale, based on the bottom-left
panel of fig.~\ref{fig:Pr-0.5} one would certainly conclude that for intermediate energies a new and 
non-universal ``mixed'' level statistics is found.

\begin{figure}
\includegraphics[width=0.42\textwidth]{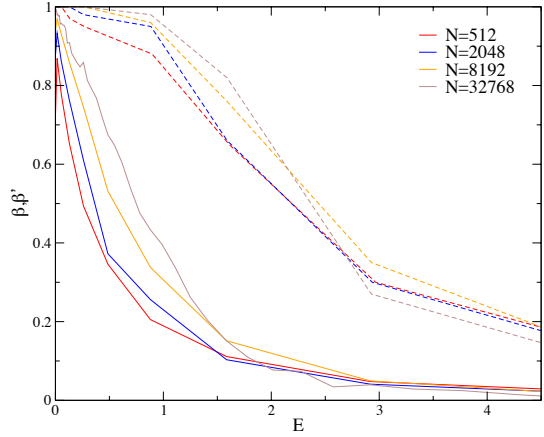}%
\caption{\label{fig:IPR-0.5} Exponents $\beta$ (continuous lines) and $\beta^\prime$ (dashed lines) describing the 
scaling with $N$ of the typical value of the Inverse Participation Ratio and of the
support set as a function of the energy for $\mu = 0.5$.}
\end{figure}

In fig.~\ref{fig:IPR-0.5} we plot the energy dependence of the 
exponents $\beta = \langle \ln \Upsilon_{2,n} \rangle_\nu/\ln N$ 
and $\beta^\prime
= \ln \langle S_\epsilon^{(n)} \rangle_\nu / \ln N$ 
describing the scaling with the system size of the typical value
of the IPR and of the average support set respectively, for $\mu = 0.5$. 
The behavior of $\beta$ and $\beta^\prime$ is coherent with previous results, at least for sufficiently small
and sufficiently large energies.
More precisely, one observes that, at fixed $N$, $\beta$ and $\beta^\prime$ decrease as the energy is increased.
Nevertheless, at fixed and small enough energy, they both grow with $N$
and seem to approach the standard value $1$ for $N \to \infty$. Conversely, 
at fixed and large enough energy, in the localized regime, $\beta$ and $\beta^\prime$
decrease to zero as the system size is increased, 
implying that $\langle \Upsilon_{2,n} \rangle_\nu,
\langle S_\epsilon^{(n)} \rangle_\nu \to \textrm{cst}$.
As mentioned above, the support set provides a more precise measure of wave-function
ergodicity compared to the IPR. 
In particular, 
the exponent $\beta$ is much smaller than one already very far from the localization transition.
In the crossover region one should expect that $\beta$ and $\beta^\prime$ show a non-monotonic behavior
as a function of $N$
on the crossover scale $N_m(E)$. However, our numerical data are too noisy to capture this behavior.
In fact, numerics based solely on the IPRs
are inconclusive and could be undoubtedly misinterpreted, 
especially for intermediate energies within the crossover regime,  
since they are affected by strong finite size effects.

\section{Numerical solution of the self-consistent equation for $Q(G)$}

In this section we provide more details on the numerical solutions of the self-consistent equation on the
probability distribution of the diagonal elements of the resolvent matrix.

In order to device an accurate and efficient algorithm to compute $Q(G)$ it is convenient to
make use, again, of the ``sparse-like'' character of LMs~\cite{metz-SM}.
As explained above, each row (or column) of $\mathcal{H}$ has $O(N)$ elements of $O(N^{-1/\mu})$ and $O(1)$ elements of $O(1)$.
We thus introduce a cut-off $\gamma$ that separates large matrix elements ($|h_{ij}|>\gamma$) from small ones ($|h_{ij}|<\gamma$).
The backbone of large entries constitutes thus a sparse (Erd\"os-R\'enyi) RM, with average
connectivity $c_\gamma = 2 N \int_\gamma^\infty P(h) \, \textrm{d}h  = \gamma^{-\mu}$~\cite{metz-SM,levy-mezard}. Since the probability distribution of small
matrix elements has now a finite variance $\sigma_\gamma^2 = 2 \int_{(N \mu)^{-1/\mu}}^\gamma h^2 P(h) \, \textrm{d}h = 
\mu \gamma^{2 - \mu}/[N(2 - \mu)]$,
their contribution to the sum in Eq.~(1) of the main text can be handled using the (classical) central limit theorem,
yielding the following self-consistent equation for $Q_\gamma (G)$ (in the limit $\gamma \to 0$):
\begin{equation} \label{eq:PG}
Q_\gamma(G) = \sum_{k=0}^\infty p_\gamma (k) \int \prod_{i=1}^k \left[ \textrm{d} G_i Q_\gamma (G_i) \, \textrm{d} h_i P(h_i) \right] 
\delta \left ( G^{-1} - E + i \eta + \sigma_\gamma^2 \langle G \rangle + \sum_{i=0}^k h_i^2 G_i \right) \, ,
\end{equation}
where $p_\gamma(k) = e^{- c_\gamma} c_\gamma^k/k!$ is the Poisson distribution of the connectivity.
This equation can be efficiently solved using a population dynamics algorithm~\cite{PD}. 
We have used a population of $2^{26}$ elements, and computed $Q_\gamma(G)$ for
$\gamma = 10^{-3},10^{-4},5 \cdot 10^{-5}$, and extrapolated the results for $\gamma \to 0$.

\begin{figure}
\includegraphics[width=0.42\textwidth]{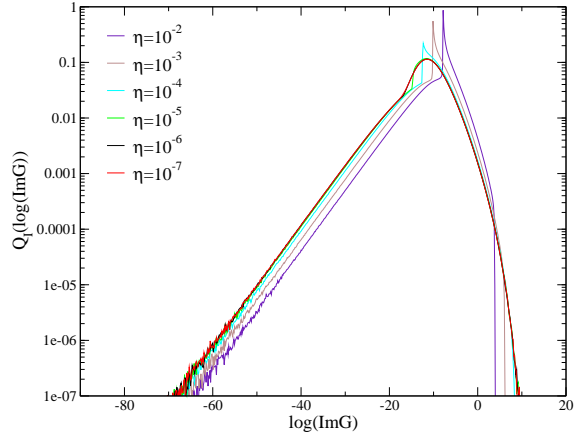}%
\caption{\label{fig:QImG5} Marginal probability distribution of $\ln \Im G$ for different values of the imaginary
regulator $\eta$ and for $E=1.25$, showing convergence to a stationary $\eta$-independent distribution
for small enough $\eta$.}
\end{figure}

In fig.~\ref{fig:QImG5} we show the marginal probability distribution of $\ln \Im G_{ii}$ for several values of the
imaginary regulator $\eta$ and for $E=20$, deep in the GOE ergodic phase.
Since the system is delocalized and the spectrum is absolutely continuous, $\tilde{Q}_I (\ln \Im G)$ must have a non-singular
limit as $\eta \to 0^+$. We indeed observe a stationary $\eta$-independent distribution for $\eta$ sufficiently
small ($\eta \lesssim 10^{-6}$). As a consequence, $\langle \Upsilon_2 \rangle \to 0$ for $\eta \to 0^+$.

\begin{figure}
\includegraphics[width=0.42\textwidth]{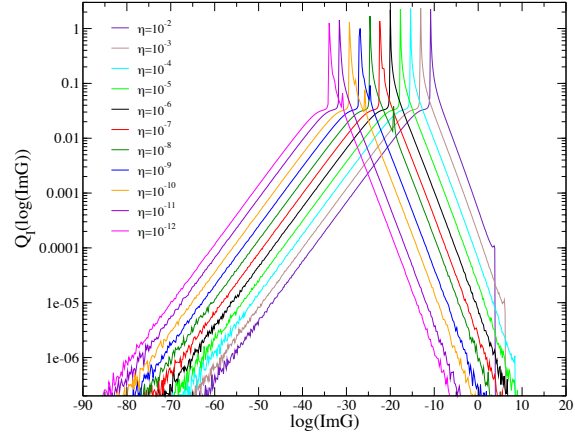}%
\caption{\label{fig:QImG22} Marginal probability distribution of $\ln \Im G$ for different values of the imaginary
regulator $\eta$ and for $E=5.5$. In the localized phase the limit $\eta \to 0^+$ is singular:
Almost all values of $\Im G_{ii}$ are of order $\eta$, except extremely rare events
described by heavy power-law tails with an exponent $1+m = 3/2$ whose coefficient vanishes as $\sqrt{\eta}$.}
\end{figure}

Conversely, in the localized phase the marginal probability distribution of the imaginary part of $G_{ii}$ has a singular
behavior as $\eta \to 0^+$, as
illustrated in fig.~\ref{fig:QImG22}. Almost all values of $\Im G_{ii}$ are of order $\eta$, except extremely rare events---whose
fraction vanishes as $\eta$---described by heavy power-law tails with an exponent $1+m$ and $m=1/2$.
More precisely,
$\tilde{Q}_I (\Im G)$ has a scaling form $f(x/\eta)\eta$ for $x \sim \eta$,
with $\int \! f(y) \, \textrm{d} y = 1$, and fat tails
\begin{equation} \label{eq:tails}
\tilde{Q}_I (\Im G) \simeq \frac{c \, \eta^{1-m}}{(\Im G)^{1+m}} \, ,
\end{equation}
with $c$ being a constant of $O(1)$,
and a cut-off for $\Im G_{ii} \simeq 1/\eta$.
Such tails gives a contribution of $O(1)$ to the density of states, whereas the bulk part only yields a vanishing
contribution.
The marginal probability distribution of the real part of $G_{ii}$ (not shown) converges to a stationary distribution with
power-law tails with a $E$-independent exponent $1+m=2$. This implies that in the localized phase 
$\langle \Upsilon_2 \rangle \to \textrm{cst}$
for $\eta \to 0^+$.

\begin{figure}
\includegraphics[width=0.43\textwidth]{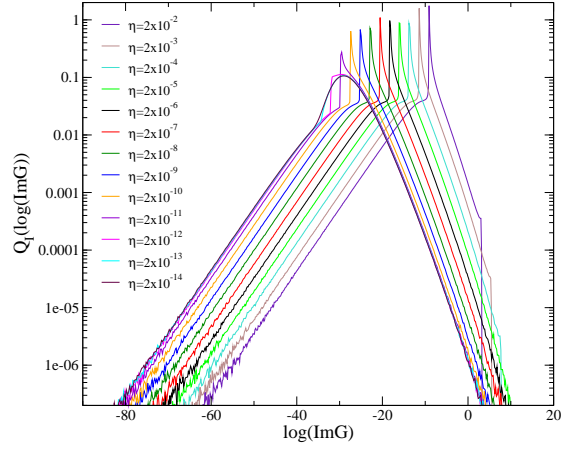}%
\caption{\label{fig:QImG13} Marginal probability distribution of $\ln \Im G$ for different values of the imaginary
regulator $\eta$ and for $E=3.25$, in the crossover phase.
For $\eta > 10^{-13}$ the system behaves as it was localized and non ergodic, showing ``singular'' power-law tails
with an exponent $1 + m$ with $1/2 \lesssim m < 1$ and a cut-off in $\Im G = 1/\eta$.
Convergence to a stationary non-singular distribution is achieved for $\eta < 10^{-13}$. }
\end{figure}

In fig.~\ref{fig:QImG13} we show the behavior of $\tilde{Q}_I (\ln \Im G)$ in the crossover region.
Since the system is delocalized, we know that the $\eta \to 0^+$ limit exists and is non-singular.
However, convergence to a stationary distribution is observed only for extremely small values of the imaginary regulator,
$\eta \lesssim 10^{-13}$ in this case.
For $\eta$ small enough but still larger than $10^{-13}$, one observes that, similarly to the
localized regime, the marginal
distribution of $\Im G_{ii}$ displays ``singular'' power-law tails described by
$\tilde{Q}_I (\Im G) \sim \eta^{1-m}/(\Im G)^{1+m}$ with
an exponent $1/2 \lesssim m < 1$, and a cut-off for $\Im G_{ii} \simeq 1/\eta$
(the exponent is instead $m \simeq 1$ for the marginal distribution of $\re G_{ii}$).
This implies that for large enough $\eta$  the tails of $\tilde{Q}_I (\ln \Im G)$ give a $O(1)$ contribution to the density of states,
whereas the bulk part gives a contribution of $O(\eta)$, as if the system was non-ergodic.

\begin{figure}
\includegraphics[width=0.42\textwidth]{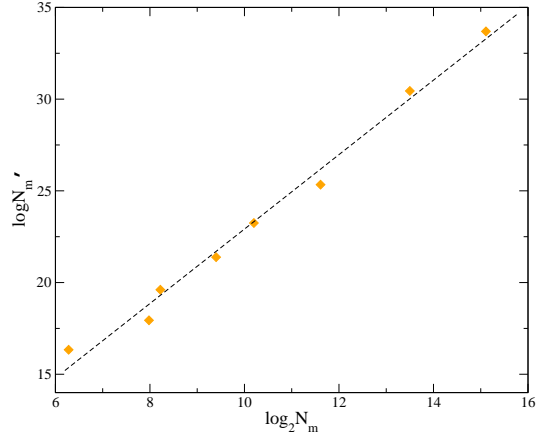}%
\caption{\label{fig:Nprime} $\ln N_m^\prime (E)$ as a function of $\ln N_m(E)$.}
\end{figure}

In order to extract the crossover scale $N_m^\prime (E)$, we measure the typical value of the imaginary part of
$G_{ii}$, $\Im G_{ii}^{typ} = e^{\langle \ln \Im G_{ii} \rangle}$, over the stationary distribution on the localized phase and,
following the argument presented in the main text, we define $N_m^\prime (E) = 1/ (\Im G_{ii}^{typ} \rho (E))$.
In fig.~\ref{fig:Nprime} we plot $\ln N_m^\prime (E)$ as a function of $\ln N_m (E)$, showing 
a linear relation between these two quantities. This implies that our argument allows to capture correctly
the origin of finite size effects.

\begin{thebibliography}{99}
\bibitem{footnote-edge} For the edge the result is different: 
one needs $\mu>4$ ~\cite{giulioJP,benarous2} as a variant of this argument would also suggest.
\end{thebibliography}
\end{document}